\documentstyle[11pt,epsf]{article}
\begin{document}
\title{Vortices in Quantum Spin Systems}
\author{John Schliemann and Franz G. Mertens\\
{\it Physikalisches Institut, Universit{\"a}t Bayreuth, D-95440 Bayreuth,
Germany}}
\date{November 1998}
\maketitle
\begin{abstract} 
%\baselineskip
We examine spin vortices in ferromagnetic quantum Heisenberg models with
planar anisotropy on two--dimensional lattices. The symmetry properties
and the time evolution of vortices built up from spin--coherent states
are studied in detail. Although these states show a dispersion typical
for wave packets, important features of classical vortices are conserved. 
Moreover, the results on symmetry properties provide a construction scheme 
for vortex--like excitations from exact eigenstates, which have a 
well--controlled time evolution. 
Our approach works for arbitrary spin length both on triangular and square 
lattices.  
\end{abstract}

%%%%%%%%%%%%%%%%%%%%%%%%%%%%%%%%%%%%%%%%%%%%%%%%%%%%%%%%%%%%%%%%%%%%%%

\section{Introduction}

Vortices are a central issue in classical models for two--dimensional
magnets, for a review see \cite{KIK:90}.
The dynamics of individual vortices has been studied extensively
for Heisenberg models with easy--plane symmetry, 
usually combining simulations performed on discrete lattices with 
analytical approaches via continuum approximations 
\cite{GWBM:89,MSB:97,KoPa:98}.
These studies have led to remarkable insight in the dynamics of
vortices in certain classical magnetic systems in terms of collective 
variables.\\
With respect to statistical properties of such systems,
vortices play a crucial role in the scenario of phase transitions of
the Kosterlitz--Thouless type, where vortices and antivortices are
bound in pairs below a transition temperature $T_{KT}$ while they
unbind above $T_{KT}$ \cite{KoTh:73}.
Following these considerations, the high--temperature phase of a planar 
ferromagnet was described by a dilute gas of topological defects, and 
the dynamic form factor of such a system was obtained using 
further reasonable approximations \cite{MBWK:89}. These results are in 
qualitative agreement with neutron scattering measurements on suitable
quasi--two--dimensional
magnetic materials. In particular, the dependence of the dynamic form factor
on wavelength and temperature is found to be consistent in
theory and experiment.
These findings strongly support the classical
description of such magnetic systems and in particular the existence of
vortex--like excitations, although important aspects of this approach are
still under discussion; for a critical overview on recent research see 
\cite{CCL:98}.\\
On the other hand, real magnetic materials consist of quantum spins.
Therefore, the question naturally arises whether quantum states exist which
mirror the essential features of classical vortices. The standard answer
given in the above literature and in many other papers is as follows:
Magnetic systems with spin lengths $S>1$ should be well described by
classical models, while for smaller spin length quantum effects become
important. Nevertheless, classical models are sometimes used also in this case,
where quantum effects are approximated by renormalizations of coupling 
parameters in the Hamiltonian, see in particular \cite{CTVV:95} and
references therein.
In this work we present a concept of quantum vortices
which is closely related to the classical limit, but takes into account
the full quantum mechanics.\\ 
The plan of this paper is as follows: In the next section we introduce the
spin model we are dealing with and summarize some of its important properties.
In section \ref{3} we examine spin vortices built
up from spin--coherent states. The results obtained there will lead us in 
section \ref{4} to a construction of vortex--like states from 
eigenstates of the Hamiltonian.

%%%%%%%%%%%%%%%%%%%%%%%%%%%%%%%%%%%%%%%%%%%%%%%%%%%%%%%%%%%%%%%%%%%%%%%%%%%%

\section{The model}

We consider a Heisenberg ferromagnet with planar exchange anisotropy 
acting on  spins of length $S$ on either a triangular or square lattice,
\begin{equation}
H=-\frac{J}{2}\sum_{<i,j>}\left[\hat S^{x}_{i}\hat S^{x}_{j}
+\hat S^{y}_{i}\hat S^{y}_{j}
+\left(1-\lambda\right)\hat S^{z}_{i}\hat S^{z}_{j}\right]\,,
\label{defmod}
\end{equation}
with $J>0$ and the sum going over all pairs of nearest neighbors. We will be 
interested in planar quantum spin vortices, whose classical counterparts are 
known to be stable for sufficiently large anisotropy parameters $\lambda$  
\cite{GWBM:89}. In the following we will always assume $\lambda=1$, which
lies in the region of classical stabilty for both lattice types. 
To construct vortices
we consider finite samples of a triangular or square lattice with open
boundaries, which have a rotational symmetry with respect to an axis 
intersecting a central plaquette; examples are shown in figure \ref{fig1}.
For definiteness and brevity we concentrate on the triangular case 
throughout this paper and only briefly comment the case of the square 
lattice, where analogous results hold.\\
For further reference let us briefly summarize some simple properties
of such systems using obvious notation. 
The Hamiltonian is invariant under rotation of all spins around the 
$z$--direction in spin space, under reversal of the $z$--component of
all spins and under appropriate rotations of the lattice. An adequate
basis of the Hilbert space is obvious: For $N$ spins 
of length $S$ we define a typical eigenstate of the $z$--component of
the total spin by $|S^{z}\rangle=\bigotimes_{i=0}^{N-1}|S^{z}_{i}\rangle$
with $S^{z}=\sum_{i}S^{z}_{i}$. The corresponding symmetry--adapted 
 basis vectors are given by  
\begin{equation}
|S^{z},m\rangle={\cal N}\left(|S^{z}\rangle
+e^{-\imath\frac{2\pi}{3}m}R|S^{z}\rangle
+e^{-\imath\frac{4\pi}{3}m}R^{2}|S^{z}\rangle\right)
\label{bs}
\end{equation}
with $m\in\{-1,0,1\}$, $\cal N$ being a normalization factor and $R$ the 
operator of a clockwise rotation of the lattice by $2\pi/3$ or, equivalently, 
a counterclockwise cyclic permutation of the local spin states. The 
states $|S^{z},m\rangle$ form invariant subspaces of (\ref{defmod}), where the 
quantum  numbers $S^{z}$, $m$ correspond to the symmetry of the model under 
rotations in spin space and real space, respectively. 
For $S^{z}\neq 0$ eigenstates of the Hamiltonian (\ref{defmod}) with
energy $E$ are denoted by 
$|S^{z},m,E\rangle$ and chosen to fullfill
\begin{equation}
|-S^{z},m,E\rangle=F|S^{z},m,E\rangle\,,
\label{flip1}
\end{equation}
where $F=\prod_{i}F_{i}$ is the spin flip operator which acts on each
lattice site as $F_{i}:|S^{z}_{i}\rangle\mapsto|-S^{z}_{i}\rangle$, or
equivalently
\begin{equation}
F\hat S_{i}^{\pm}F^{+}=\hat S_{i}^{\mp}
\quad,\quad F\hat S_{i}^{z}F^{+}=-\hat S_{i}^{z}\,.
\label{flip2}
\end{equation} 
Note that $F$ is the same as a rotation of the spins by $\pi$ around the
$x$--axis up to a possible phase factor; namely it holds 
$\exp(\pm\imath\pi\hat S^{x}_{i})=(\pm\imath)^{2S}F_{i}$. For $S^{z}=0$ 
eigenstates can be characterized further by the spin flip symmetry and are
denoted by $|0,m,E,f\rangle$ with $f\in\{-1,1\}$ being the eigenvalue of $F$.
Generally, each eigenstate with quantum numbers
$S^{z}$, $m$ has got degenerate counterparts in subspaces with the same
values of $|S^{z}|$, $|m|$. Degenerate eigenstates which differ only in
the sign of $m$ are related by a complex conjugation of the spin wave 
function.\\
The case of a square lattice is obviously analogous; one simply has to 
infer in equation (\ref{bs}) rotations by $\pi/2$ instead of $2\pi/3$ with 
$m\in\{-1,0,1,2,\}\pmod{4}$.

%%%%%%%%%%%%%%%%%%%%%%%%%%%%%%%%%%%%%%%%%%%%%%%%%%%%%%%%%%%%%%%%%%%%%%

\section{Vortices built out of spin--coherent states}
\label{3}

We now examine vortices which are built up from spin--coherent states on each
lattice site. Such objects have recently been discussed by the present authors
within a semiclassical approach \cite{ScMe:98}. Here we take into account the
full quantum mechanics.\\
In the Hilbert space of a spin of length $S$ a spin--coherent 
state $|S;\vartheta,\varphi\rangle$ is defined by the equation
\begin{equation}
\vec s_{\vartheta,\varphi}\cdot\hat{\vec S}\,|S;\vartheta,\varphi\rangle
=\hbar S\,|S;\vartheta,\varphi\rangle
\label{def1}
\end{equation}
for the direction $\vec s_{\vartheta,\varphi}=
(\sin\vartheta\cos\varphi,\sin\vartheta\sin\varphi,\cos\vartheta)$ 
\cite{Rad:71}.
These states can be considered as the immediate quantum analogue to classical
spin vectors. In the usual basis they can be expressed as
\begin{equation} 
|S;\vartheta,\varphi\rangle=U(\vartheta,\varphi)\,|S\rangle=
\sum_{n=0}^{2S}{2S\choose n}^{\frac{1}{2}}
\left(\cos\left(\frac{\vartheta}{2}\right)\right)^{2S-n}
\left(\sin\left(\frac{\vartheta}{2}\right)\right)^{n}
e^{\imath\varphi(n-S)}\,|S-n\rangle
\label{defcs}
\end{equation}
with 
\begin{equation}
U(\vartheta,\varphi)=\exp\left(-\frac{\imath}{\hbar}\varphi\hat S^z\right)
\exp\left(-\frac{\imath}{\hbar}\vartheta\hat S^y\right)\,.
\label{defU}
\end{equation}
For our purposes we shall define here the vorticity $\nu$ of a quantum state 
completely analogously to the classical case by
\begin{equation}
\nu=\sum_{i\rightarrow j}\Delta\varphi_{i,j}\qquad,\qquad
\Delta\varphi_{i,j}=\left(\varphi_{j}-\varphi_{i}\right)\in[-\pi,\pi[\,.
\label{defvor}
\end{equation} 
The sum is taken over a closed path on the lattice in counterclockwise 
direction and the classical--like angles $\varphi_{i}$ are given by local
expectation values of spin operators,
\begin{equation}
\varphi_{i}=\tan^{-1}\frac{\langle\hat S^{y}_{i}\rangle}
{\langle\hat S^{x}_{i}\rangle}\pmod{2\pi}\,.
\label{defphi}
\end{equation}
Thus, the vorticity is a nonlinear functional of the underlying 
quantum state.
In the following we shall restrict ourselves to the case $|\nu|=1$.\\
We now model a planar vortex as a tensor product of spin--coherent states,
\begin{equation}
|\psi_{\pm}\rangle=\bigotimes_{i=0}^{N-1}|S;\vartheta_{i},\varphi_{i}\rangle\,,
\label{csv}
\end{equation}
where we take $\vartheta_{i}=\pi/2$ for all $i$ and the angles $\varphi_{i}$
to be given by the classical values as depicted in the examples of figure
\ref{fig1}. This choice leads to a vortex (denoted by $|\psi_{+}\rangle$ with
$\nu=1$), while the mapping $\varphi_{i}\mapsto-\varphi_{i}$ converts
 it into an 
antivortex $|\psi_{-}\rangle$ with $\nu=-1$. From (\ref{defcs}) one sees that 
$|\psi_{+}\rangle$ and $|\psi_{-}\rangle$ are related via a complex
conjugation of the spin wave function, which is the same here as a spin flip
\begin{equation}
|\psi_{+}\rangle=F|\psi_{-}\rangle\,.
\end{equation}    
Before examining further the quantum states (\ref{csv}) let us briefly 
remark on the classical vortex. As seen in figure \ref{fig1}, in the small
system a) all directions of the classical spins can be derived by intuitive
symmetry arguments and are the same as in the well--known continuum solution
$\varphi(x,y)=\tan^{-1}(y/x)+{\rm constant}$ with $x$, $y$ denoting cartesian
coordinates in the plane. In the system b) the same holds 
for the inner lattice sites labelled by 0, 1, 2 and 5, 8, 11, but not for the
outer ones. E.~g. it is easy to see that the sum of the classical vectors
on 3 and 4 must be always parallel to the spin on 0, but the exact value of, 
say,
$\phi_{3}$ must be calculated in detail and turns out to be different from the
continuum solution. In system c) of figure \ref{fig1} one has the same 
situation for the sites 4, 5 and so on.
Note also that the classical vortex is a static solution only if
its center coincides with the center of the system, because otherwise
its image vortices created by the boundaries cause a movement of the vortex
center. Within the continuum approximation of the system this situation
is completely analogous to two--dimensional electrostatics.\\
Now we analyze the states $|\psi_{\pm}\rangle$ with respect to the 
symmetries of the Hamiltonian. Let us concentrate again on the triangular case.
The scalar product of the vortex (\ref{csv}) with a typical basis vector 
(\ref{bs}) is
\begin{equation}
\langle S^{z},m|\psi_{\pm}\rangle={\cal N}\left(
\langle S^{z}|\psi_{\pm}\rangle
+e^{+\imath\frac{2\pi}{3}m}\langle S^{z}|R^{+}|\psi_{\pm}\rangle
+e^{+\imath\frac{4\pi}{3}m}
\langle S^{z}|(R^{+})^{2}|\psi_{\pm}\rangle\right)\,.
\end{equation}
The application of $R^{+}=R^{-1}$ on $|\psi_{\pm}\rangle$ is the same as
a counterclockwise (clockwise) rotation of each local spin--coherent 
state by an angle of $2\pi/3$, i.~e. all angles $\varphi_{i}$ in 
(\ref{csv}) get a turn of $\pm 2\pi/3$. Therefore, with the help of 
(\ref{defcs}) one finds for integer spin length $S$
\begin{equation}
\langle S^{z},m|\psi_{\pm}\rangle\Bigg\{
{\propto\langle S^{z}|\psi_{\pm}\rangle\qquad S^{z}\mp m=0\pmod{3}
\atop{=0\qquad\qquad\qquad{\rm otherwise\qquad}\nonumber}}
\label{sr1}
\end{equation}
and similarly for half--integer $S$
\begin{equation}
\langle S^{z},m|\psi_{\pm}\rangle\Bigg\{
{\propto\langle S^{z}|\psi_{\pm}\rangle\qquad S^{z}+\frac{N}{2}\mp m=0\pmod{3}
\atop{=0\qquad\qquad\qquad{\rm otherwise\qquad\qquad}\nonumber}}\,.
\label{sr2}
\end{equation}
These relations determine the invariant subspaces of the Hamiltonian
in which the vortex state $|\psi_{\pm}\rangle$ has non--vanishing overlap. 
We therefore call them selection rules. To cover the case of the square 
lattice one simply has to replace (mod 3) with (mod 4).\\
The square moduli of the coefficients in the expansion (\ref{defcs})
form a binomial distribution of range $2S$ with a probability parameter
$p=\sin^{2}(\vartheta/2)$. This is the probability distribution for the
results of measurements of the $z$--component of an individual spin being
in the state (\ref{defcs}). By an elementary theorem of stochastics,
the distribution of a finite sum of quantities which are binomial--distributed
with a common parameter $p$ is again of the binomial type with the 
same parameter and a range just given by the sum of the individual ranges.
In a planar vortex we have $p_{i}=\sin^{2}(\pi/4)=1/2$ for all $i$ and 
therefore
\begin{equation}
\sum_{m}\sum_{E}|\langle S^{z},m,E|\psi_{\pm}\rangle|^{2}
={2SN\choose SN-S^{z}}\frac{1}{2^{2SN}}\,.
\label{bindis}
\end{equation}
The sum goes over all eigenstates having $S^{z}$ as the quantum
number of the total spin; for $S^{z}=0$ the states $|0,m,E,f\rangle$ have to be
inserted and summed over $f$ as well. The mean value of this symmetric 
distribution is of course zero and the square variance is given by
\begin{equation}
\left(\Delta S^{z}\right)^{2}=\frac{SN}{2}\,.
\label{width}
\end{equation}
According to the central limit theorem of stochastics, the expression 
(\ref{bindis}) approaches a Gaussian shape for large $SN$, where
\begin{equation}
\lim_{SN\to\infty}
\sum_{a\Delta S^{z}\leq S^{z}\leq b\Delta S^{z}}
\sum_{m,E}|\langle S^{z},m,E|\psi_{\pm}\rangle|^{2}
=\int_{a}^{b}dx\,g(x)
\label{gauss1}
\end{equation}
with the normalized Gaussian distribution $g(x)=1/\sqrt{2\pi}\exp(-x^{2}/2)$ 
and real numbers $a<b$. If one fixes a certain value of $m$ in the above
summations, only every third value of $S^{z}$ gives a non--vanishing
contribution because of the selection rules (\ref{sr1}), (\ref{sr2}). Thus
we find
\begin{equation}
\lim_{SN\to\infty}\sum_{S^{z},E}|\langle S^{z},m,E|\psi_{\pm}\rangle|^{2}
=\int_{-\infty}^{+\infty}dx\,g(3x)=\frac{1}{3}\,,
\label{gauss2}
\end{equation}
i.~e. for an infinite system, $N\to\infty$, or in the classical limit,
\begin{equation}
\hbar\to 0\quad,\quad S\to\infty\quad,\quad\hbar S={\rm constant}\,,
\label{clim}
\end{equation}
the (anti--)vortex $|\psi_{\pm}\rangle$ has the same square overlap in all 
subspaces characterized by different rotational quantum numbers 
$m\in\{-1,0,1\}$.\\
The same arguments hold for the square lattice with $3$ to be replaced
with $4$ in (\ref{gauss2}).\\
In summary, the above equations characterize the states $|\psi_{\pm}\rangle$
with respect to the symmetries of the system. In figure \ref{fig2} we have
illustrated the results for the system shown in figure \ref{fig1}a)
($N=6$) and $S=5/2$.\\\\
Next we analyze the states $|\psi_{\pm}\rangle$ with respect to the
exact eigenstates of the model (\ref{defmod}). To this end we have 
numerically diagonalized the full Hamiltonian for small systems, i.~e. have
computed {\em all} eigenvalues and eigenvectors in the invariant subspaces.
This procedure can be done with today's computers for the system a) in 
figure \ref{fig1} for spin lengths $S=1/2,1,\dots,5/2$, while for larger 
lattices like b) and c) one is still restricted to $S=1/2$.\\
Let us first consider the system shown in figure \ref{fig1}a). 
The expectation
value of the Hamiltonian is 
\begin{equation}
\langle\psi_{\pm}|H|\psi_{\pm}\rangle=-\frac{3}{2}J(\hbar S)^{2}\,.
\label{erg}
\end{equation}
Its variance has been obtained in reference \cite{ScMe:98} and reads here
\begin{equation}
\Delta H=J(\hbar S)^{2}\frac{3}{4S}\,.
\end{equation}
As it must be, this quantity vanishes in the classical limit (\ref{clim}).
Figure \ref{fig3} shows histograms of the square overlap of 
$|\psi_{\pm}\rangle$ with the eigenstates of the Hamiltonian and the density 
of states for $S=2$ as a function of the energy; for other spin length
$S=1/2,1,\dots,5/2$ the data looks qualitatively similar.
The time evolution of the system being initially in the state 
$|\psi_{\pm}\rangle$ can be followed in terms of
 \begin{equation}
\langle\psi_{\pm}(0)|\psi_{\pm}(t)\rangle
=\langle\psi_{\pm}|e^{-\frac{\imath}{\hbar}Ht}|\psi_{\pm}\rangle\,,
\label{sp}
\end{equation}
which is essentially the Fourier transform of the data shown in the upper
diagram of figure \ref{fig3}. Therefore this quantity decays on a
time scale given by the uncertainty relation
\begin{equation}
\Delta H\Delta t\geq\frac{\hbar}{2}\,,
\label{urel}
\end{equation}
which is in full agreement with our numerical findings  even for 
comparatively large spin lengths like $S=5/2$, where a classical--like
behaviour of spin systems is often assumed. Thus, even for large $S$
the time dependence of the scalar product
(\ref{sp}) is the same as for any usual dispersive wave packet, in contrast 
to the classical vortex which is a nonlinear coherent
excitation. This is due to the fact that the classical limit (\ref{clim}) is 
{\em not} approached properly by taking a large spin length but keeping 
$\hbar$ as finite as it is.\\
Inspite of this general statement, the analysis of local expectation values
\begin{equation}
\langle S_{i}^{\alpha}(t)\rangle
:=\langle\psi_{\pm}|e^{\frac{\imath}{\hbar}Ht}\hat S_{i}^{\alpha}
e^{-\frac{\imath}{\hbar}Ht}|\psi_{\pm}\rangle
\end{equation}
shows that certain features of the initial vortex structure nevertheless
 remain 
present in the time evolution of the state. First note that the local spin
expectation values on sites $i$, $j$ connected by a rotation of the  lattice,
i.~e. $\hat S_{j}^{\alpha}=R\hat S_{i}^{\alpha}R^{+}$, are related by 
\begin{equation}
\langle S_{j}^{z}(t)\rangle=\langle S_{i}^{z}(t)\rangle\qquad,\qquad
\langle S_{j}^{+}(t)\rangle=e^{\pm\imath\frac{2\pi}{3}}
\langle S_{i}^{+}(t)\rangle\,,
\label{rot1}
\end{equation}
as it is intuitively obvious and can be shown easily with the help of 
(\ref{sr1}), (\ref{sr2}). Therefore, the vorticity of the central plaquette
is necessarily conserved.\\
Let us now discuss the time evolution of the vortex in more detail.
For the system in figure \ref{fig1}a) our numerical results are as follows. 
The expectation values $\langle S_{i}^{z}(t)\rangle$ are strictly zero 
for all times, lattice sites and spin lengths $S=1/2,1,\dots,5/2$. 
This is surprising since only the 
$z$--component of the {\em total} spin is conserved due to symmetry. In figure
\ref{fig4} we have plotted the time evolution of the in--plane spin 
components for the state $|\psi_{+}\rangle$ and $S=2$.
The upper diagram shows 
$|\langle S_{i}^{+}(t)\rangle|$, which may be seen as an `effective 
spin length'. This quantity decays for both classes of
lattice sites on a time scale given by (\ref{urel})
to comparatively small numbers and even becomes zero for certain times. 
The lower diagram shows the direction angles $\varphi_{i}$ calculated from 
(\ref{defphi}). Surprisingly, these angles remain constant up to
changes of $\pi$, which occur when $|\langle S_{i}^{+}(t)\rangle|$ goes
through zero, i.~e. the spin expressed by its expectation values reverses
its direction. The times when such reversals occur are not
identical for both classes of sites, but apparently strongly correlated.
Note that this gives rise to quantum fluctuations of the vorticity as 
defined in (\ref{defvor}). For instance, if the inner lattice sites 0, 1, 2
have undergone such a reversal while the outer ones have not, the vorticity
on the three outer plaquettes is changed from $0$ to $-1$, while the
vorticity of the central plaquette is preserved as mentioned before. 
An evaluation of the first $1000$ time units after starting the dynamics shows 
that in about $80\%$ of this time intervall the vorticities on all plaquettes
have their initial values, while in the remaining time the vorticities
of the outer plaquettes are changed to $\pm 1$. This shows the strong 
correlation in the spin dynamics also seen in figure \ref{fig4}. It is an
interesting speculation whether such fluctuation phenomena are related to the 
sponteneous appearance of vortex--antivortex pairs (in larger systems), which
is well--known from classical spin models.\\
The findings described above hold similarly for all spin lengths 
$S=1/2,1,\dots,5/2$ and both types of states $|\psi_{\pm}\rangle$.\\
In the system of figure \ref{fig1}b) some new observations are made.
As already mentioned the numerical analysis is restricted to $S=1/2$.
The spins on the inner lattice sites 0, 1, 2 and 5, 8, 11 show completely
the same behaviour as in the system described before, while the time
evolution of spins on the outer sites, say 3 and 4, is different. Here
small $z$--components $\langle S_{i}^{z}(t)\rangle$ arise which
are plotted in the upper diagram of figure \ref{fig5}. For symmetry
reasons these quantities differ in sign on sites which are inequivalent
under rotation, since the expectation value of the $z$--component ot the total
spin is constantly zero. Moreover, also the in--plane angles $\varphi_{i}$
are not conserved (up to changes by $\pi$) as shown in the lower diagram.
But remarkably, the sum 
$\langle\vec S_{3}(t)\rangle+\langle\vec S_{4}(t)\rangle$ is always
parallel or antiparallel to $\langle\vec S_{0}(t)\rangle$ with the 
orientations being correlated in a similar way as described before. 
This is also a strong reminiscence of the classical vortex structure.\\
The differences in the behaviour of the spins found in the system of
figure \ref{fig1}b) are a surprising parallel to our previous remark on
spin directions in the classical vortex. Here the outer lattice sites
are also distinguished from the inner ones, since their spins are not described
by the static continuum solution.\\
Summarizing, we have demonstrated that the states $|\psi_{\pm}\rangle$,
although they show dispersion, preserve
important properties of classical vortices.

%%%%%%%%%%%%%%%%%%%%%%%%%%%%%%%%%%%%%%%%%%%%%%%%%%%%%%%%%%%%%%%%%%%%%%

\section{Vortices constructed from exact eigenstates}
\label{4} 

Since the time dependence of the spin vortices presented in the last section is
rather complicated, it is desirable to find vortex--like quantum states
which have a well--controlled time evolution. To this end the symmetry
rules (\ref{sr1}), (\ref{sr2}) provide a simple construction scheme.
It is useful to distinguish three different cases:\\\\
\underline{({\it i})} Triangular lattice, $SN$ integer: A vortex--like
quantum state is given in terms of exact eigenstates of the Hamiltonian 
by the following {\em ansatz}:
\begin{equation}
|\chi_{\pm}\rangle=\frac{1}
{\sqrt{|\alpha_{-1}|^{2}+|\alpha_{0}|^{2}+|\alpha_{1}|^{2}}}
\Big(\alpha_{-1}|-1,\mp 1,E_{1}\rangle
+\alpha_{0}|0,0,E_{0},f\rangle+\alpha_{1}|1,\pm 1,E_{1}\rangle\Big)
\label{chi1}
\end{equation}
This is a linear combination of eigenstates which has nonzero amplitudes only
for quantum numbers `allowed' by the rules (\ref{sr1}), (\ref{sr2}) and is 
restricted to the most important contributions with $|S^{z}|\leq 1$. From
each invariant subspace only one eigenstate is involved.
The states with $S^{z}=\pm 1$ are chosen to be degenerate 
(which is always possible). Thus (\ref{chi1}) is effectively a 
two--level--system
with an internal frequency $\omega=(E_{1}-E_{0})/\hbar$. Denoting 
$|\chi_{\pm}(t)\rangle=\exp(-(\imath/\hbar)Ht)|\chi_{\pm}\rangle$
one finds similarly as in (\ref{rot1}):
\begin{equation}
\langle\chi_{\pm}(t)|\hat S_{j}^{+}|\chi_{\pm}(t)\rangle
=e^{\pm\imath\frac{2\pi}{3}}
\langle\chi_{\pm}(t)|\hat S_{i}^{+}|\chi_{\pm}(t)\rangle\,,
\label{rot2}
\end{equation}
where the lattice sites $i$, $j$ are related by a rotation. Therefore the 
central plaquette carries a constant vorticity of $\nu=\pm 1$. For
$|\alpha_{-1}|=|\alpha_{1}|$ it holds:
\begin{equation}
\langle\chi_{\pm}(t)|\hat S_{i}^{z}|\chi_{\pm}(t)\rangle\equiv 0
\label{ex1}
\end{equation}
for all sites $i$ and the expectation values of the in--plane components
are given by
\begin{eqnarray}
\langle\chi_{\pm}(t)|\hat S_{i}^{+}|\chi_{\pm}(t)\rangle
& = & \frac{2|\alpha_{1}\alpha_{0}|\langle 0,0,E_{0},1|\hat S_{i}^{+}
|1,\pm 1,E_{1}\rangle}{2|\alpha_{1}|^{2}+|\alpha_{0}|^{2}}
\nonumber\\
& & \cdot \exp\left(\frac{\imath}{2}\left(\phi_{-1}-\phi_{+1}\right)\right)
\cos\left(\omega t+\phi_{0}-\frac{1}{2}\left(\phi_{-1}+\phi_{+1}\right)\right)
\,,
\label{ex2}
\end{eqnarray}
where we have inferred $\alpha_{l}=|\alpha_{l}|\exp(\imath\phi_{l})$ and
assumed $f=1$ for simplicity; the case $f=-1$ leads only to unimportant
additional phase factors. To derive (\ref{ex1}), (\ref{ex2}) the relations
(\ref{flip1}), (\ref{flip2}) have been used. \\
Thus, the vector of the local expectation values of the spin components 
has a constant direction (up to reversal) on each site while its
length varies harmonically with the frequency $\omega$ in time.
Differently from the vortices constructed from spin--coherent states,
all vectors of expectation values lie strictly in the plane and
their reversals, i.~e. the zeros of their length, occur simultaneously
on all lattice sites. Therefore no fluctuations of vorticity arise.\\
The above construction works on triangular lattices of the type shown
in figure \ref{fig1} and of arbitrary size. It provides quantum
states with a very simple time evolution and typical properties of vortices.\\
To illustrate this, let us return to the system 1a). Figure \ref{fig6} shows
the lower part of the spectrum for $S=2$ as a function of $S^{z}$. The
ground state has quantum numbers $S^{z}=0$, $f=0$ and is part of a band
of states with $m=0$. Well separated from this we have a band of degenerate 
states with $m=\pm 1$ and next a more or less continuum--like distribution of 
states. This turns out to be qualitatively the same for the other spin lengths
considered here. The choice of states to be used in (\ref{chi1}) is certainly
not unique. To give a definite example, let us choose the combination
with lowest possible energy, i.~e. we take the state with $S^{z}=0$, $m=0$
to be the ground state and the other ones from the excited elementary band. 
Note that the expectation value of energy for such a linear combination is 
much lower than the expression (\ref{erg}), which corresponds directly to 
the energy of a classical vortex. 
\begin{table}
\begin{center}
\begin{tabular}{|c|c|c|c|c|}
\hline
$S$ & $|\langle\chi_{+}|\hat S_{0}^{+}|\chi_{+}\rangle|$ & 
$\varphi_{0}$ & $|\langle\chi_{+}|\hat S_{3}^{+}|\chi_{+}\rangle|$ &
$\varphi_{3}$ \\
\hline
1/2 & 0.0522 & $0^{\circ}$ & 0.1785 & $60^{\circ}$ \\
1 & 0.0675 & $0^{\circ}$ & 0.2646 & $60^{\circ}$ \\
3/2 & 0.0814 & $0^{\circ}$ & 0.3295 & $60^{\circ}$ \\
2 & 0.0938 & $0^{\circ}$ & 0.3835 & $60^{\circ}$ \\
\hline
\end{tabular}
\end{center}
\caption{Expectation values of in--plane spin components in the state
$|\chi_{+}\rangle$ constructed from elementary excitations (see text)
\label{tab1}}
\end{table}
In table \ref{tab1} we present the expectation values
of the in--plane spin components for different spin lengths in the state 
$|\chi_{+}\rangle$ at time $t=0$,
where we have set all $|\alpha_{l}|=1$ and adjusted the phases $\phi_{l}$ in a 
manner that the argument
of the cosine in (\ref{ex2}) vanishes and the spin on the site $0$
has $\varphi_{0}=0$. Obviously, the `effective spin lengths' are
strongly reduced compared with the original ones. This was also found in the
previous section in the time evolution of a vortex built out of spin--coherent
states (cf. figure \ref{fig4}). The importance of this effect has also
been stressed by other authors using a variational approach to the
spin dynamics \cite{Ost:86,IKS:92}.\\
As mentioned above, the central plaquette has
vorticity $\nu=1$, and the directions on the other sites also exactly
mirror the classical vortex structure. This is a non--trivial property since
the only strict relation between these directions is given by (\ref{rot2}).
This observation can also be made for other choices of eigenstates, mainly from
the lower part of the spectrum, but for an arbitrary linear combination
of the form (\ref{chi1}) this is not the case.\\
Thus we have demonstrated the existence of vortex--like quantum states 
built up from elementary excitations , whose energy is much lower
than the semiclassical vortex examined in the previous section.
The energy of a single (semi--)classical vortex is known to grow
logarithmically with the size of the system. The construction
presented here relies only on symmetry properties and works 
for arbitrary system size. Therefore, the energy of the vortex--like quantum 
state discussed in the above example must be assumed to remain in the order
of the exchange integral $J$ (or at least finite)
even in an infinite system, since otherwise the energy difference
between the lowest and the first excited band of quantum states would
have to grow with increasing system size
to macroscopic values (or infinity), which is completely unlikely.\\\\
We only sketch the remaining two cases .\\\\
\underline{({\it ii})} Triangular lattice, $SN$ half--integer: Here all
values of $S^{z}$ are also half--integer and a vortex--like quantum state
can be constructed as (cf. (\ref{sr2})):
\begin{equation}
|\chi_{\pm}\rangle=\frac{1}
{\sqrt{|\alpha_{-1}|^{2}+|\alpha_{1}|^{2}}}
\left(\alpha_{-1}|-\frac{1}{2},\mp 1,E\rangle
+\alpha_{1}|\frac{1}{2},\pm 1,E\rangle\right)\,,
\label{chi2}
\end{equation}
which also has the properties (\ref{rot2}), (\ref{ex1}) for 
$|\alpha_{-1}|=|\alpha_{1}|$. Therefore the spin
structure expressed in local expectation values is also planar 
and the central plaquette carries a vorticity of $\pm 1$. The spin structure
on other plaquettes depends on further details and can be examined as above.
As the involved eigenstates with different quantum numbers are chosen 
degenerate,
we obtain an {\em exact} eigenstate which has typical features of a vortex,
at least with respect to its center.\\\\
\underline{({\it iii})} Square lattice, $SN$ necessarily integer: This case
is merely analogous to ({\it i}) with the extension that the eigenstate
in (\ref{chi1}) with $S^{z}=0$ may may be chosen from two
different subspaces ($m=0$ or $m=2$). This slightly generalizes the
selection rules
(\ref{sr1}), (\ref{sr2}), which give only one of these two possibilities, 
and leads also to different vorticities.

%%%%%%%%%%%%%%%%%%%%%%%%%%%%%%%%%%%%%%%%%%%%%%%%%%%%%%%%%%%%%%%%%%%%%%

\section{Conclusions}
\label{5}

\begin{figure}
\begin{center}
\epsfysize=8cm
\epsffile{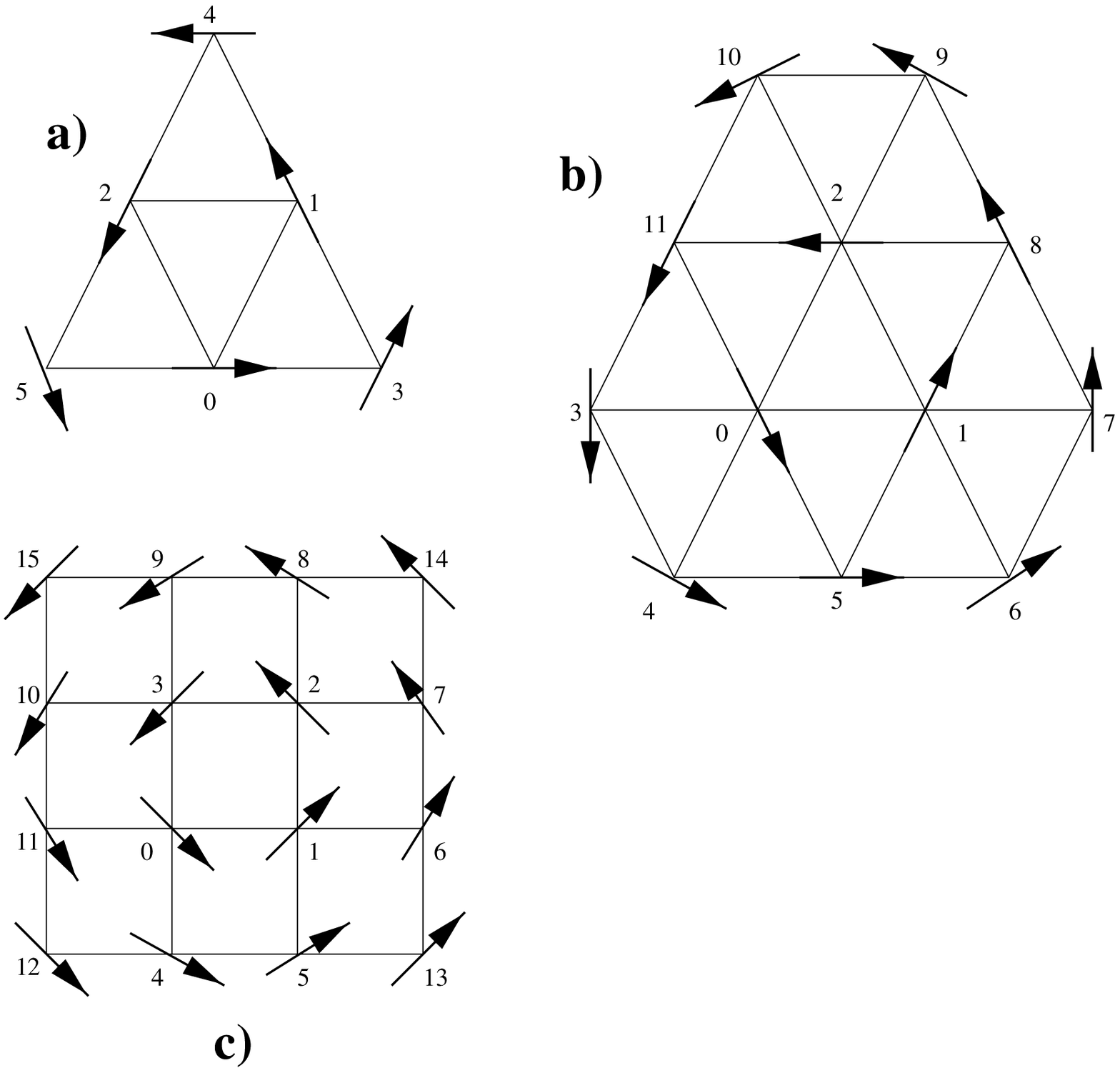}
\end{center}
\caption{Classical vortex structrures on lattice samples which have a 
rotational symmetry axis intersecting the center of the vortex\label{fig1}}
\end{figure}

\begin{figure}
\begin{center}
\epsfysize=8cm
\epsffile{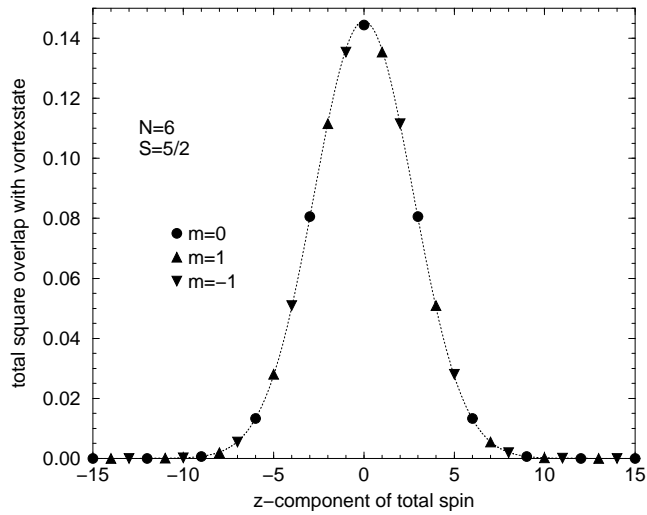}
\end{center}
\caption{Total square overlap of $|\psi_{+}\rangle$ in invariant subspaces
of the Hamiltonian. The dashed line is a Gaussian with width $\Delta S^{z}$.
\label{fig2}}
\end{figure}

\begin{figure}
\begin{center}
\epsfysize=8cm
\epsffile{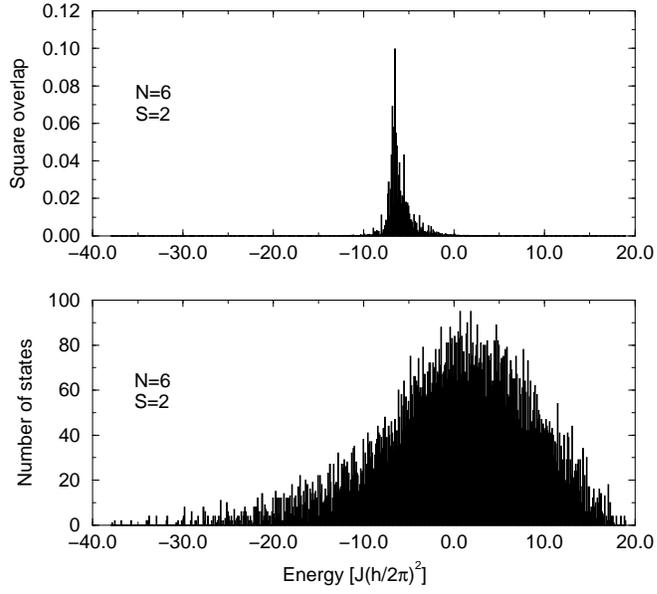}
\end{center}
\caption{Square overlap of $|\psi_{\pm}\rangle$ with eigenstates of the 
Hamiltonian (upper diagram) and density of states as a function of the energy
for the system of figure 1a) and $S=2$
\label{fig3}}
\end{figure}

\begin{figure}
\begin{center}
\epsfysize=8cm
\epsffile{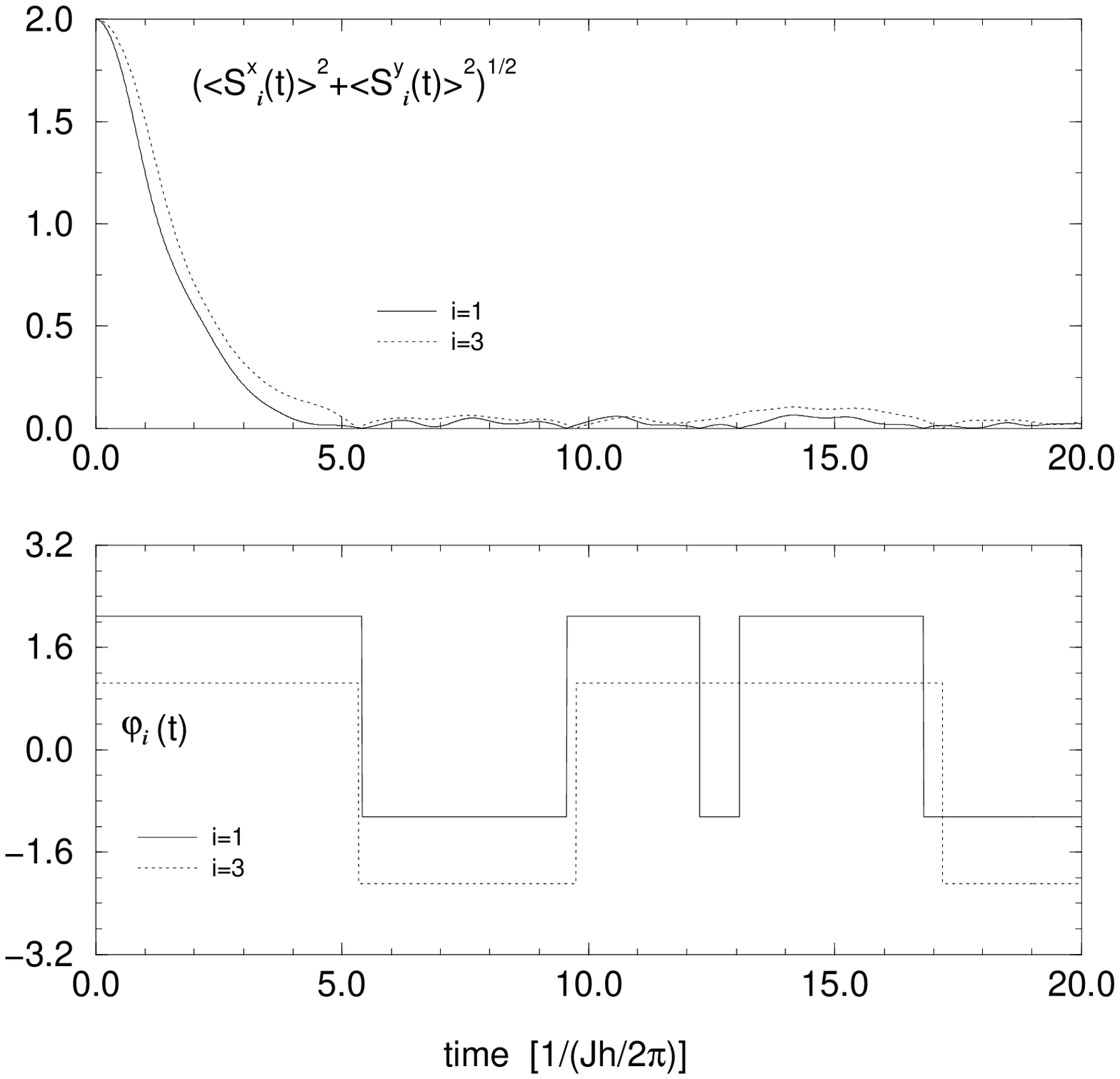}
\end{center}
\caption{Time evolution of local expectation values in the state 
$|\psi_{+}\rangle$ for the system of figure 1a) and $S=2$
\label{fig4}}
\end{figure}

\begin{figure}
\begin{center}
\epsfysize=8cm
\epsffile{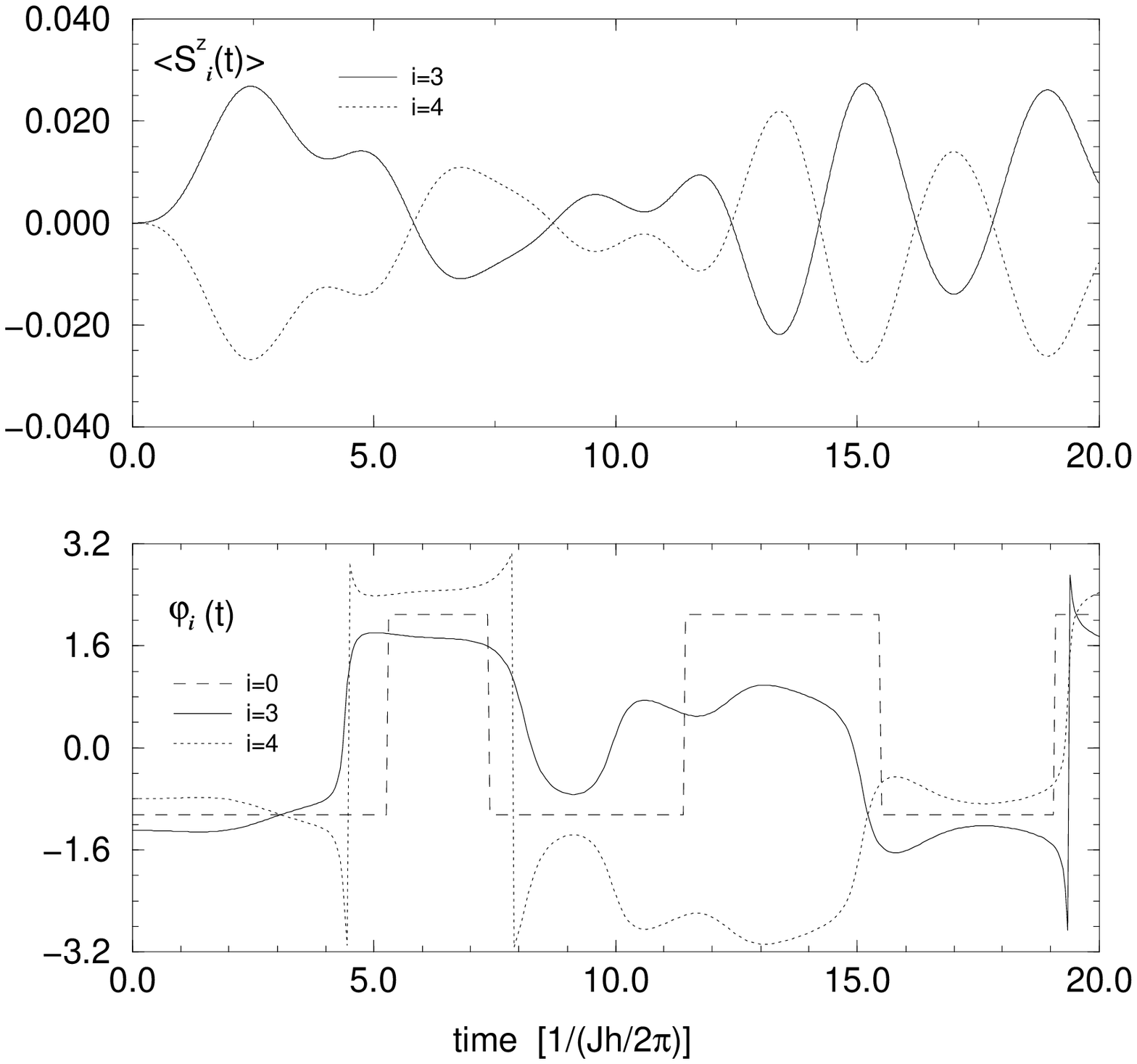}
\end{center}
\caption{Time evolution of local expectation values in the state 
$|\psi_{+}\rangle$ for the system of figure 1b) and $S=1/2$
\label{fig5}}
\end{figure}

\begin{figure}
\begin{center}
\epsfysize=8cm
\epsffile{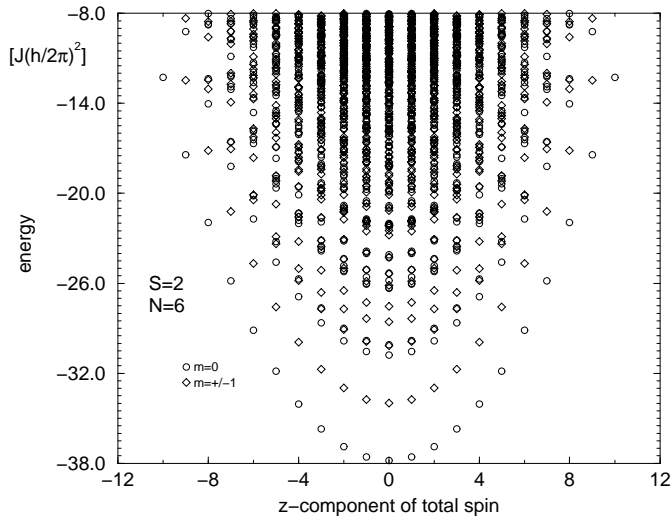}
\end{center}
\caption{The low--lying part of the spectrum of the system of figure 1a)
for $S=2$
\label{fig6}}
\end{figure}

In this work we have examined planar quantum spin vortices in ferromagnetic
Heisenberg models taking into account the full quantum mechanics.\\
Vortices built up from spin--coherent states are studied in detail.
These objects can be seen as the immediate quantum analogue of
a classical static vortex on a discrete lattice.
Important results on their symmetry properties are given by the relations
(\ref{sr1}), (\ref{sr2}), (\ref{bindis}) and are illustrated in figure
\ref{fig2}. The time evolution of such vortices is in general quite complicated
and, from a global point of view, typical for quantum mechanical wave packets.
On the other hand, a detailed numerical study of the local spin expectation
values shows that important properties of the initial classical--like
vortex structure are conserved. This may be viewed as a reminiscence
of the topological character of the classical vortex solution,
although such topological arguments do not apply strictly in a discrete
system.\\
Our symmetry analysis heavily relies on the symmetry of the underlying 
lattice sample
with respect to the vortex center as shown in figure \ref{fig1}. To
characterize a vortex whose center lies not in the center of system,
one may consider a subsystem which has this property. The quantum vortex
state projected onto the Hilbert space of this subsystem should have similar 
properties as found here, e.~g. the selection rules 
(\ref{sr1}), (\ref{sr2}) should hold approximately, but not exactly.
We expect the deviations from these selection rules to result in a movement
of the vortex center, as it is well--known from the classical vortex.\\ 
Our observations should generally raise the confidence in applying the
classical Kosterlitz--Thouless theory to real magnetic systems
(consisting of quantum spins) in the spirit of an effective field theory
and with respect to critical behavior, where many details of the system
can be expected to be unimportant.\\
The symmetry properties of such vortices lead to a `reduced' construction
of vortex--like excitations in terms of exact eigenstates of the Hamiltonian 
as described in the foregoing section. We obtain vortex--like quantum states
involving only two different energy levels or, in particular cases, exact
eigenstates having vortex--like features.
Moreover, we find such states, which have unambigously the properties of
a classical vortex and a very simple time evolution, even at energies
which are much lower than the classical vortex energy. This may, on the other
hand, indicate some important modification of the role of vortices in
the quantum system compared with the classical case.
Concerning this issue further investigations are desirable,
which may possibly start from the construction scheme of vortex--like
excitations in terms of exact eigenstates given in this work.\\
The approach presented here is expected to be useful also for other cases
like non--planar vortices in ferromagnets or vortices in antiferromagnets.\\\\

{\bf Acknowledgement:} The authors are gratefull to Alexander Wei{\ss}e and
Gerhard Wellein for friendly help with computer details, and to Frank
G{\"o}hmann for a critical reading of the manuscript. F.~G.~M. would like
to thank J. Zittartz for drawing his attention to the fact that only little
was known so far about vortices in quantum spin systems.\\
This work has been supported by the Deutsche Forschungsgemeinschaft under
grant No. Me534/6-1.
 
%%%%%%%%%%%%%%%%%%%%%%%%%%%%%%%%%%%%%%%%%%%%%%%%%%%%%%%%%%%%%%%%%%%%%%

\newpage

\end{document}